# Getting the Right Twist: Influence of Donor–Acceptor Dihedral Angle on Exciton Kinetics and Singlet–Triplet Gap in Deep Blue Thermally Activated Delayed Fluorescence Emitter

*Sebastian Weissenseel, ‡ Nikita A. Drigo,† Liudmila G. Kudriashova,‡ , Markus Schmid, ∇ Thomas Morgenstern,∇ Kun-Han Lin,§ Antonio Prlj,$, § Clémence Corminboeuf,§ Andreas Sperlich*,‡ Wolfgang Brütting,∇ Mohammad Khaja Nazeeruddin,*,† and Vladimir Dyakonov,‡*

‡ Experimental Physics VI, Julius-Maximilian University of Würzburg, 97074 Würzburg, Germany

† Group for Molecular Engineering of Functional Materials, École Polytechnique Fédérale de Lausanne (EPFL), CH-1951 Sion, Switzerland

∇ Experimental Physics IV, University of Augsburg, 86159 Augsburg, Germany

§ Laboratory for Computational Molecular Design, EPFL, CH-1015 Lausanne, Switzerland

$ Laboratory of Theoretical Physical Chemistry, EPFL, CH-1015 Lausanne, Switzerland

**ABSTRACT:** Here, a novel deep blue emitter SBABz4 for use in organic light-emitting diodes (OLED) is investigated. The molecular design of the emitter enables thermally activated delayed fluorescence (TADF), which we examine by temperature-dependent time-resolved electroluminescence (trEL) and photoluminescence (trPL). We show that the dihedral angle between donor and acceptor strongly affects the oscillator strength of the charge transfer state alongside the singlet–triplet gap. The angular dependence of the singlet–triplet gap is calculated by time-dependent density functional theory (TD-DFT). A gap of 15 meV is calculated for the relaxed ground state configuration of SBABz4 with a dihedral angle between the donor and acceptor moieties of 86°. Surprisingly, an experimentally obtained energy gap of 72±5 meV can only be explained by torsion angles in the range of 70°−75°. Molecular dynamics (MD) simulations showed that SBABz4 evaporated at high temperature acquires a distribution of torsion angles, which immediately leads to the experimentally obtained energy gap. Moreover, the emitter orientation anisotropy in a host matrix shows an 80% ratio of horizontally oriented dipoles, which is highly desirable for efficient light outcoupling. Understanding intramolecular donor-acceptor geometry in evaporated films is crucial for OLED applications, because it affects oscillator strength and TADF efficiency.

1. Introduction

Fluorescence-based organic light-emitting diodes (OLEDs) are known to reach a maximum internal quantum efficiency (IQE) $\eta_{int}$ of 25% due to spin-statistics, while the incorporation of heavy metal atoms in emitting molecules enables the phosphorescence IQE of 100% due to efficient spin–orbit coupling.[1–4] The molecules based on thermally activated delayed fluorescence (TADF) can harvest non-radiative triplets via reverse intersystem crossing (RISC) and can also yield $\eta_{int}$ of 100%.[5-9] Intuitively, one has to minimize the singlet–triplet energy gap in order to enhance the triplet-to-singlet up-conversion.[10,11] This condition is commonly fulfilled by separating electron and hole wave functions onto different moieties, so that their overlap is reduced. Large separation of electron and hole is realized in the molecules comprised of donor and acceptor units that are almost perpendicular to each other. The dihedral angle Θ between donor and acceptor determines the exchange energy and hence $\Delta E_{ST}$. Actually, one has to find a compromise for Θ, because Θ = 90° means perfectly separated e–h wave functions and, consequently, zero singlet–triplet gap, but, at the same time, the oscillator strength f for radiative transitions from the excited CT state to the ground state vanishes. [12] $\Theta$ can be calculated theoretically, but the optimum value for $\Theta$ is hard to predict a priori. The calculated dihedral angle $\Theta$ for an isolated molecule in its relaxed configuration can differ from the real value in a thin-film device prepared by thermal evaporation. Therefore, the experimentally measured activation energies can deviate substantially from the theoretically predicted values. Thus, the relation between experimentally found activation energies and $\Delta E_{ST}$ is not evident. However, molecular dynamics simulation (MD) can reveal distribution of dihedral angles in evaporated films of TADF emitters, thereby delivering activation energies comparable with experimental values.

In order to make use of the increased IQE, the light outcoupling should be optimized, too. The light outcoupling in OLEDs with isotropic emitters can be quantified by the outcoupling factor $\eta_{out} = 1/(2n_2)$, where $n$ is the refractive index of the emitting layer of the device.[13] The molecular orientation in evaporated amorphous films for OLEDs is in most cases random. This leads to an isotropic light-emission that limits external quantum efficiency (EQE) of OLEDs. However, recent studies have shown that sticklike emitters tend to self-orient themselves within the film.[14–16] The resulting anisotropic emission improves the light-outcoupling, which results in the higher theoretical limit and experimental values for EQE. Optimizing the light extraction of OLEDs is, therefore, a highly relevant aspect.

In this work, we demonstrate an increased horizontal orientation of the transition dipole moment in a film of a deep blue TADF emitter. The emitter can harvest the non-radiative triplets and also yield an $\eta_{int}$ close to 100%. In addition to that, we performed optical simulations on the OLED stack with a dipole model based on the scattering-matrix-transfer-formalism in order to calculate the maximum possible EQE in dependence of the film thicknesses in the OLED device. However, we were able to reveal the actual distribution of dihedral angles in the evaporated film of TADF emitters using molecular dynamics simulation (MD). The resulting theoretical values for activation energies show excellent agreement with experiment.

The deep blue emitting molecule studied here consists of two donor (D) and two acceptor (A) moieties. The central core consists of two acridine donors linked by a spiro-bridge (x). Together with the two linearly attached benzonitrile acceptors, an A-DxD-A compound is realized. The molecular structure 4,4'-(10H,10'H-9,9'-spirobi[acridine]-10,10'-diyl)dibenzonitrile (SBABz4) is shown in **Figure 1a**. The spiro bridge leads to orthogonal arrangement of acridine.[17,18] The H-atoms of the acceptor force the dihedral angle between the benzonitrile and acridine to open up.[19] This nearly perpendicular conformation between the donor and acceptor decreases the orbital overlap of the highest occupied molecular orbital (HOMO) and the lowest unoccupied molecular orbital (LUMO), which will result in the desired small singlet–triplet gap. Moreover, the spiro-connection was reported to rigidify the molecular structure and, eventually, to stabilize the planarized acridine conformation with the perpendicular arrangement of D-A pair, thus this twisted conformation favors emission via TADF channel. [20]

With a full theoretical description of the transition rates of the three-level model system for the excited charge transfer states we are able to describe the rate of the reverse intersystem crossing $k_{RISC}$ of SBABz4 by two experimental available observables, the TADF efficiency, $\phi_{TADF}$, and the lifetime of the delayed fluorescence $\tau_{TADF}$.

2. Results and Discussion

The ground state geometry of SBABz4 was calculated by utilizing density functional theory (DFT), explained in detail in the Supporting Information. In ground state geometry, donor (acridine) and acceptor (benzonitrile) moieties are nearly orthogonal to each other (~ 86°). The Kohn–Sham HOMO and LUMO are shown in **Figure 1a**. There is a clear spatial separation between HOMO and LUMO. Since the molecule is a composition of two donor and two acceptor units, there are two nearly degenerate HOMOs and LUMOs which are additionally shown in Figure S1, together with the natural transition orbitals (NTOs) in Figure S2 and S3. The NTOs demonstrate the charge transfer character of the low-lying excited states of SBABz4 molecule. We also derived the energy of the locally-excited triplet state of the donor (denoted as $T_3$ in Table S1 in Supporting Information) and it lies energetically above the degenerate excited singlet charge-transfer $S_1$, $S_2$ (CT) states.

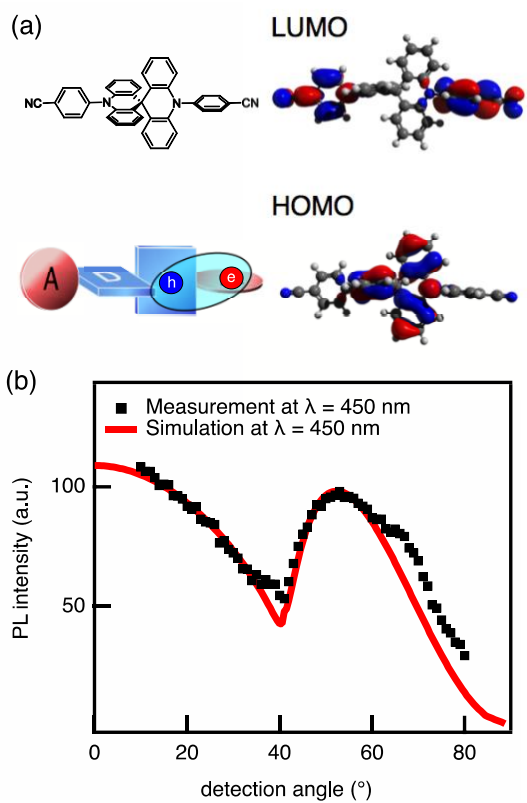

**Figure 1.** (a) left-hand side: structural formula of SBABz4 and a sketch of the linear A-DxD-A structure. right-hand side: Kohn–Sham LUMO and HOMO of SBABz4. (b) Angle-resolved PL intensity of co-evaporated 10wt%SBABz4:DPEPO at a wavelength of 450 nm (squares). The fit (red trace) reveals a fraction of 80% of horizontal dipoles in the film. Note that all emission wavelengths were taken into account in the analysis but only that for 450 nm is shown here.

In order to determine the orientation of the transition dipole moment, angular-dependent PL spectrometry as proposed by Frischeisen et al. was performed on a 20 nm thick film of SBABz4:DPEPO (10:90) co-evaporated on a glass substrate.[21] As shown in **Figure 1b**, the measurement reveals a fraction of 80% of the emitting dipoles lying within the substrate plane. This result confirms the horizontal orientation of the sticklike SBABz4 molecule as expected. Additionally, the intrinsic quantum efficiency of the emitter in the host was determined from transient PL measurements as described in more detail in the Supporting Information Figure S6. The derived value of 0.36 ± 0.07 is in good agreement with the PL quantum yield published earlier.[22] Both, the fraction of horizontal dipoles as well as the intrinsic quantum efficiency have a major impact on the device performance.[23]

In order to study the theoretical limit for the performance of the presented devices in more detail, we performed optical simulations based on the plane wave decomposition as described by Barnes et al. and Penninck et al. and described by the scattering matrix formalism.[24-26] Our software allows for the calculation of the maximum possible EQE in dependence of the film thickness and refractive index in the stack layout assuming balanced charge carrier injection and a singlet-triplet ratio of unity.

To optimize the OLED layer stack, simulations for the EQE in dependence of the NPB (hole transport layer) thickness and the TPBi (electron transport layer) thickness between 5 and 100 nm were performed (**Figure 2a**). The transport layers were chosen as they are adjustable with the least influence on the electrical properties of the device. Note that the Purcell effect modifying the intrinsic quantum efficiency of the emitter and the predominantly horizontal emitter orientation are taken into account in the simulation. The optimum device structure in the studied ranges was identified at layer thicknesses of 41 nm and 5 nm of TPBi and NPB, respectively, yielding a possible EQE of 11.6%. The simulation shows no local maximum with respect to the NPB layer, because of relatively high thicknesses of ITO and PEDOT:PSS in the stack. For experimental reasons, we chose slightly different thicknesses (50 nm of TPBi and 20 nm of NPB) that predict the maximum possible efficiency of 10.0%, which is still very close to the optimum value.

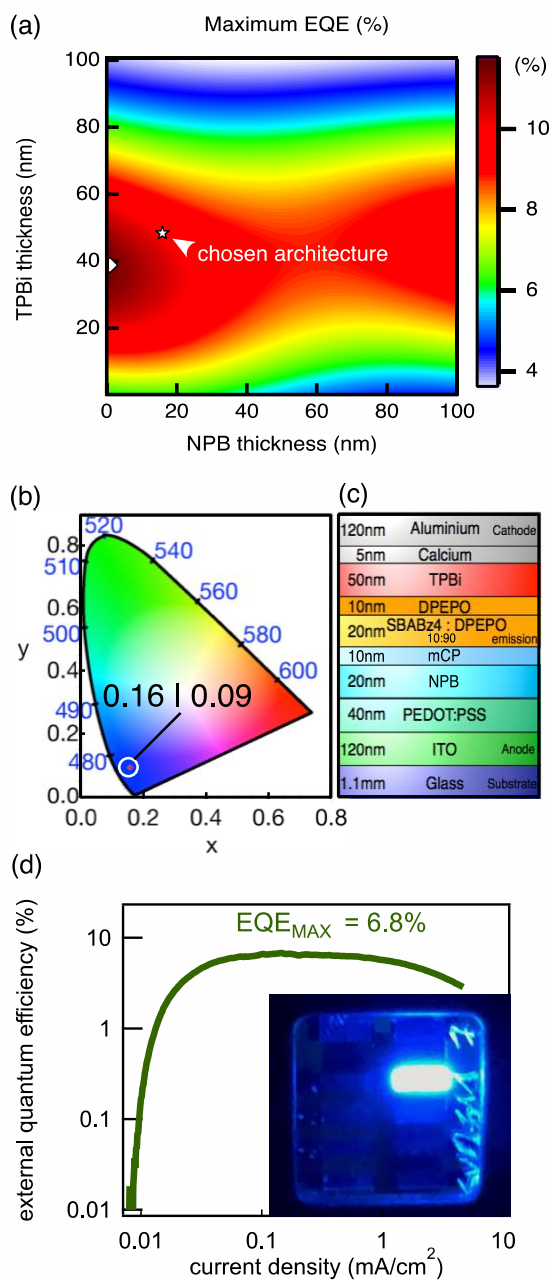

**Figure 2.** (a) Simulation of the maximum possible EQE (color-coded on the right) in dependence of the transport layer thicknesses. The marked values (open symbols) reveal that the realized device structure (star symbol) is very close to the theoretical optimum (diamond symbol). (b) CIE-color coordinates (0.16, 0.09). (c) Schematic device stack of 10wt% SBABz4:DPEPO based OLED (d) EQE of a working SBABz4:DPEPO device versus current density. The inset shows a photograph of the operating device.

We fabricated OLEDs with a 20 nm thick emitting layer of 10wt% SBABz4 doped in a matrix of bis[2-(diphenylphosphino)phenyl] ether oxide (DPEPO). The normalized photoluminescence (PL) and electroluminescence (EL) spectra are shown in Figure S7 and demonstrate deep blue emission, which is represented by the Commission Internationale de l'Éclairage (CIE) color coordinates (0.16, 0.09) of PL **(Figure 2b)**. The optimized layer structure of the device is PEDOT:PSS (40 nm) / NPB (20 nm) / mCP (10 nm) / 10%wt SBABz4:DPEPO / DPEPO (10 nm) / TPBi (50 nm) / Ca (5 nm) / Al (120 nm) **(Figure 2c)**. NPB and TPBi serve as hole and electron transport layers (PEDOT:PSS = poly(3,4-ethylenedioxythiophene)-poly(styrenesulfonate), NPB = N,N′-di(1-naphthyl)-N,N′-diphenyl-(1,1′-biphenyl)-4,4′-diamine, mCP = 1,3-bis(N-carbazolyl)benzene, TPBi = 2,2′,2"-(1,3,5-benzinetriyl)-tris(1-phenyl-1-H-benzimidazole)). The mCP and DPEPO neat layers are used as electron, hole and exciton blocking layers, respectively.[27–29] For the emission layer a co-evaporated layer of 10%SBABz4:DPEPO is included, where SBABz4 is a dopant in a DPEPO matrix. Figure S8 presents a typical *J-V*-curve of this device. A maximum EQE of 6.8% was achieved **(Figure 2d).** Comparing this value with the theoretically predicted 10 %, we clearly see potential for future device optimization. The emission from the SBABz4:DPEPO device is featureless and consistent with the PL spectrum of a pure SBABz4 film.[22] Emission from the DPEPO matrix is not observed, neither in PL nor EL. An inset of Figure 2d shows a photograph of a working device.

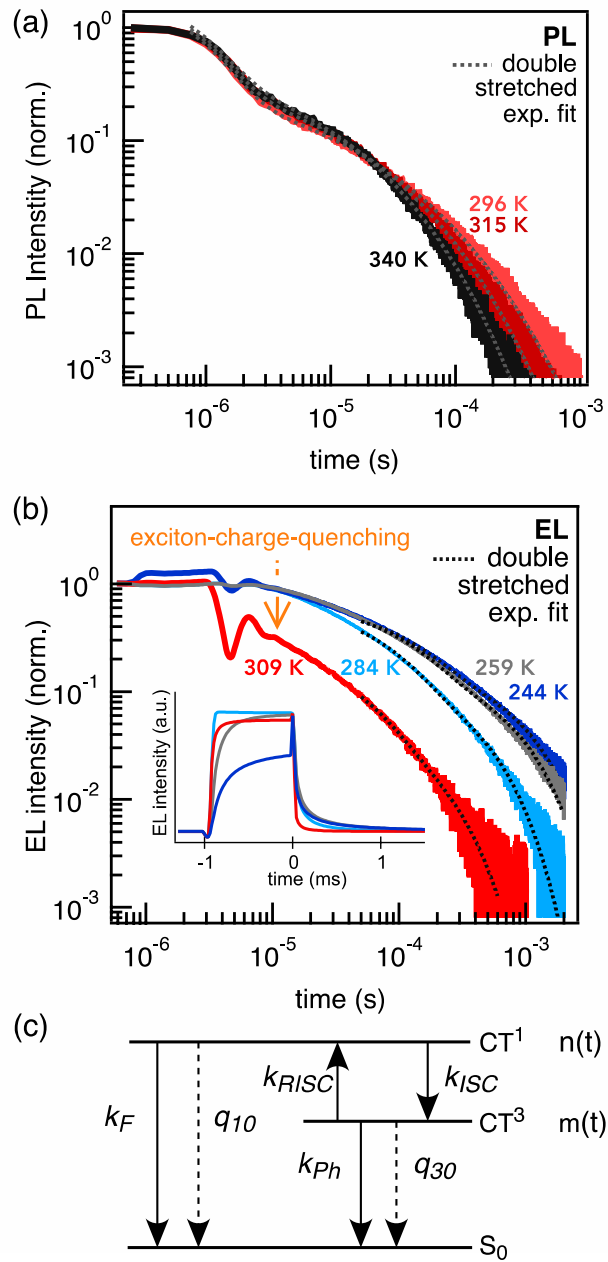

**Figure 3.** (a) trPL and (b) trEL-transients. A double stretched exponential fit is applied to extract the temperature dependence of kRISC. The inset shows the trEL transients in lin-lin-scale. (c) Rate model for a three-level-system of SBABz4. The dashed lines indicate loss mechanisms to the ground state.

**Figure 3a and 3b** show the time-resolved measurements of SBABz4:DPEPO for PL and EL. The transient curves contain prompt and delayed components, as was shown recently.[22] We extracted the average decay time of the transients from the stretched exponential fits (dotted lines). Note, that in the case of transient EL the device is turned on at -1 ms and turned off at 0 ms for trEL measurements (inset Figure 3b). At these switching points, a crosstalk between current leads and detector was unavoidable. This causes oscillating spikes but does not affect the delayed emission after $10^{-5}$ s. With increasing temperature, the delayed component shows a faster decay for both PL (Figure 3a) and EL (Figure 3b). In order to evaluate the temperature activation of the reverse intersystem crossing, we can assume a three-level system consisting of an excited CT singlet ($CT_1$) and triplet ($CT_3$) state, as well as a ground state ($S_0$) **(Figure 3c)**. The population of the $CT_1$ and $CT_3$ states are denoted as $n(t)$ and $m(t)$, respectively. The transition rate constants $k$ are marked with solid arrows, while the dashed lines indicate non-radiative loss mechanisms to the ground state ($q$). In this scheme, the rate constant of reverse intersystem crossing is marked as $k_{RISC}$. Now we can express the rate equations for n, m:

$$\frac{dn}{dt} = -(k_F + q_{10} + k_{ISC}) \cdot n + k_{RISC} \cdot m \quad (1)$$

$$\frac{dm}{dt} = k_{ISC} \cdot n - (k_{Ph} + q_{30} + k_{RISC}) \cdot m \quad (2)$$

By solving the rate equations (detailed analysis in the Supporting Information), we can express the rate constant $k_{RISC}$ by the TADF efficiency $\phi_{TADF}$ and the decay time of the delayed emission $\tau_{TADF}$ trough

$$\ln(k_{RISC}) = \ln\left(\frac{\phi_{TADF}}{\tau_{TADF}}\right) + C_1 = \frac{-E_A}{k_B T} + C_2 \quad (3)$$

$C_1$ and $C_2$ are temperature independent constants. $\phi_{TADF}$ is evaluated by integrating the delayed part of the transient over time. The resulting area is proportional to the number of detected photons. The decay time $\tau_{TADF}$ can be extracted from the double stretched exponential fit with $\alpha = 0.5$: [30]

$$I(t) = A \cdot exp\left(-\left[\frac{t}{\tau_{pr}}\right]^\alpha\right) + B \cdot exp\left(-\left[\frac{t}{\tau_{TADF}}\right]^\alpha\right) \quad (4)$$

With expression (3), we are able to determine the activation energy $E_A$ from trPL and trEL. Note that equation (3) takes into account only phonon-assisted non-radiative decay of triplets (denoted with $q_{30}$), which is valid for the contactless trPL measurement.

However, additional non-radiative processes, such as exciton–charge quenching, can take place in complete devices. The trEL data shows a clear hint for non-radiative exciton–charge recombination.

Namely, the rapid decrease of intensity at 309K (orange arrow above the red curve in Figure 3b) at $10^{-5}$ s. There, the intensity of the transient is lower compared to other temperatures. At higher temperatures, more charges are available because of a higher current density. Note that the measurements are taken at constant voltage, while the current density is increasing with temperature. Altogether, because of exciton–charge-quenching, equation (3) underestimates $k_{RISC}$ values found from trEL at high concentration of free charges, that is, at high temperatures.

The Arrhenius plot for RISC derived from temperature-dependent trPL and trEL is shown in **Figure 4a**. A linear behavior is observed for trPL from 295 K to 340 K. The activation energy according to equation (1) is (72±5) meV. This temperature activation is also valid for the EL of the device. The term $ln(k_{RISC})$ is increasing with temperature from 244 K to 264 K. With higher temperature, the device is suffering from loss mechanisms mentioned above, which are also temperature activated. Because of the enhanced losses at elevated temperatures (>264 K) the trEL shows a decrease for $ln(k_{RISC})$. These losses are absent in trPL, which leads to the assumption, that the emission layer itself is not affected by these losses.

In contrast to the experimentally determined activation energy, TD-DFT computations for the ground state conformation of SBABz4 ($\Theta$ = 86°) predict a value of $\Delta E_{ST}$ = 15 meV. If the dihedral angle is deviating from this conformation, the orbital overlap between HOMO and LUMO is increasing. Since the exchange integral $J_{ex}$ is linked to the singlet–triplet gap $\Delta E_{ST} = E_S - E_T = 2\, J_{ex}$, the gap between the singlet $E_S$ and triplet $E_T$ energy increases as well.[31]

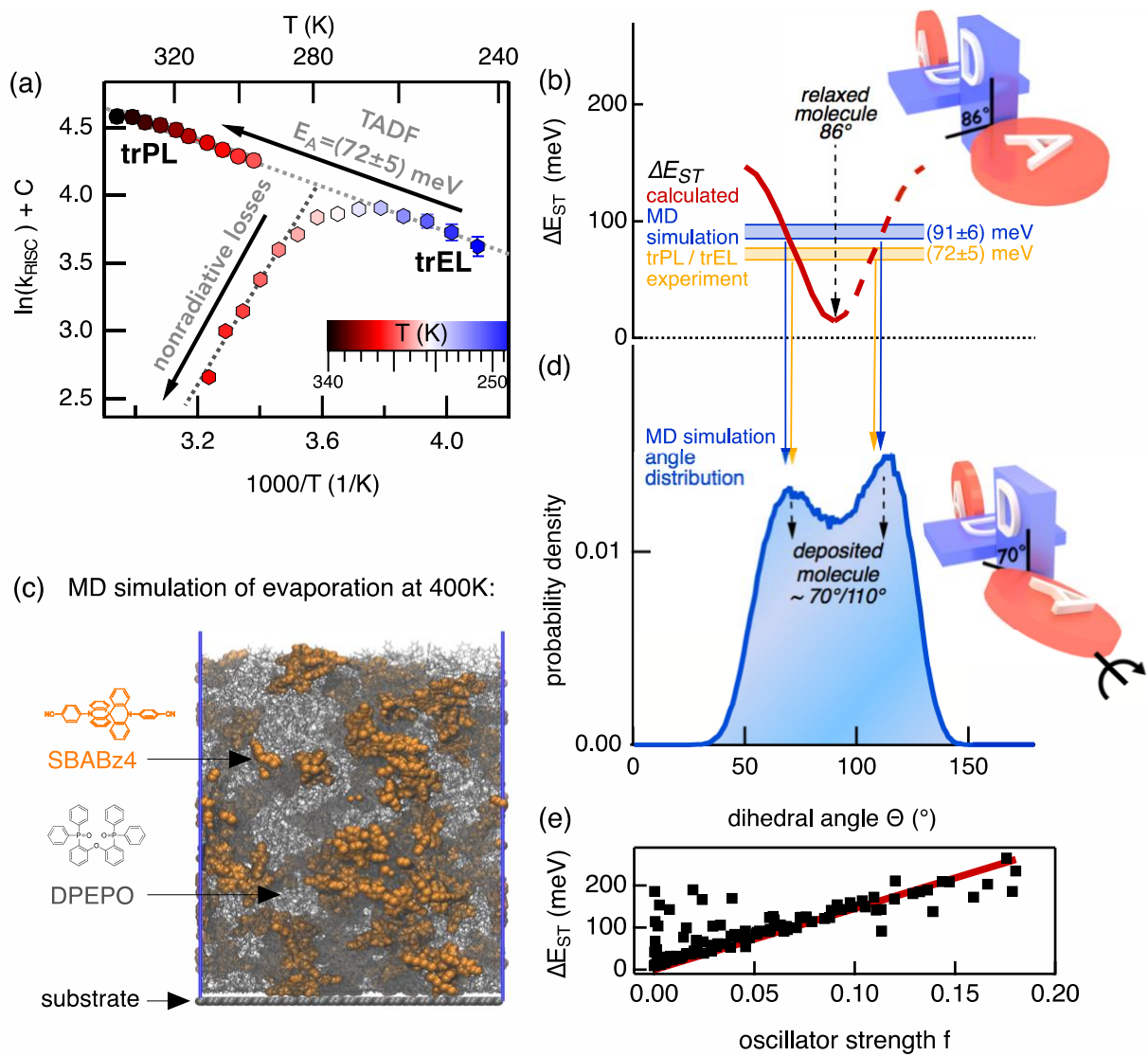

**Figure 4.** (a) From the Arrhenius plot follows an activation energy for RISC $\Delta E_A = (72 \pm 5)$ meV (b) Computed singlet–triplet gap $\Delta E_{ST}$ (red trace) for SBABz4 in dependence of the dihedral angle $\Theta$ between A and D (with the other A-D dihedral angle kept fixed at 90°). TD-DFT calculations for a relaxed molecule predict a dihedral angle $\Theta = 86°$. Experimental value for $\Delta E_{ST}$ (orange); average $\Delta E_{ST}$ obtained from molecular dynamics (MD) simulations (blue). (c) MD simulation mimicking the evaporation process for a film of 10mol% of SBABz4 (orange space-filling models) in DPEPO (grey sticklike models). For each molecule in the deposited film, TD-DFT calculations are performed to extract a mean $\Delta E_{ST}$ and the distribution of $\Theta$. (d) Probability density of $\Theta$ averaging 1000 MD simulation snapshots. (e) The oscillator strength $f$ and singlet–triplet gap $\Delta E_{ST}$ for each molecule from MD simulation (black dots). $f$ is larger for molecules with twisted donor–acceptor conformation and hence, larger singlet–triplet gap (red line: guide to the eye). All excited-state computations were performed using TDA/TD-DFT along with γ-tuned LC-BLYP/6-31G(d,p) (see SI for more details).

**Figure 4b** shows the calculated singlet–triplet gap $\Delta E_{ST}$ in dependence of the dihedral angle (red curve). The experimental value is marked in orange. This indicates that the conformation of SBABz4 in an evaporated layer is different from the relaxed molecule in vacuum. Hot evaporation ($T = 468$ K) of SBABz4 with DPEPO matrix supplies the necessary energy to twist the acceptor in the film (Figure S5). To gain more insight into the influence of the evaporation on the dihedral angle, we performed molecular dynamics (MD) simulations.[32,33] In the calculations a film of SBABz4:DPEPO with the ratio of 1:9 is generated, with 1000 molecules in total. The system is then equilibrated at 300 K **(Figure 4c)**. After reaching the equilibrium, 1000 snapshots of the film are averaged to obtain a statistical distribution of the torsion angles between D and A in all SBABz4 molecules. For each molecule in the deposited film TD-DFT computations are performed to extract a distribution of values for $\Delta E_{ST}$. To include the dielectric effect, we adopt an implicit solvent model in TD-DFT computations with the dielectric constant $\varepsilon = 3$. The mean $\Delta E_{ST}$ value, which takes into account all simulated SBABz4 molecules, is 91±6 meV (blue region in Figure 4b). The probability density of the dihedral angle in **Figure 4d** also indicates two local maxima at $\Theta_1 \approx 70°$ and $\Theta_2 \approx 110°$. These maxima of the angle distribution coincide with the experimental value obtained from trPL and trEL (marked with orange arrows) and with the theoretical value obtained from MD simulations (marked with blue arrows).

Combining the experimental value with MD simulation, we conclude that the dihedral angle of the molecule in an evaporated film is different from the relaxed value of $\Theta = 86°$. The actually dominant dihedral angles in the evaporated film have an impact on the oscillator strength $f$. **Figure 4e** shows $f$ and singlet-triplet gap $\Delta E_{ST}$ for each molecule obtained from MD simulation. SBABz4 in the relaxed conformation shows only weak emission properties. The favored conformation in the film is substantial to reach efficient emission, while the increased singlet–triplet gap also constricts the efficiency of the RISC process of SBABz4.

The observation that the trEL and trPL traces in Figures 3a, b can only be fitted by stretched exponential functions and not by ordinary exponentials is a direct result of the distribution of dihedral angles and thus singlet-triplet gaps and oscillator strengths. We expect that earlier (later) emitting molecules have larger (smaller) $\Delta E_{ST}$ and larger (smaller) $f$.

3. Conclusion

In this work, we presented a novel sticklike molecule SBABz4 with A-DxD-A structure for deep blue emitting OLED applications. It enables TADF under optical and electrical excitation. In a co-evaporated SBABz4:DPEPO films, the horizontal orientation of the emission transition dipole moment was found to be 80%, which is essential for enhanced light outcoupling. A detailed analysis of the underlying rate equations for temperature-dependent transient PL and EL revealed an activation energy of $\Delta E_{ST}$ = 72±5 meV.

The OLED device layer stack was optimized for maximum EQE by optical modelling of the transport layer thicknesses, which was reproduced experimentally. By implementing advanced molecular dynamics and TD-DFT simulations, we recreated the OLED co-evaporation process and explained the experimental results by the actual molecular conformation in the co-evaporated thin film. We can thus link the activation energy to a mean dihedral angle of $\Theta \approx 70 - 75°$ between the donor and acceptor units which deviates decisively from $\Theta = 86°$ for a relaxed molecule in vacuum. The latter would offer a lower $\Delta E_{ST}$, but would also result in a vanishing oscillator strength $f$ for radiative transitions. Thus, the optimal emitter design and deposition for the next-generation OLEDs requires finding a balance between the low activation energy for efficient TADF and sufficient oscillator strength for high total emission.

4. Experimental Section

*Material:* The SBABz4 emitter has been synthesized and purified according to the published protocol.[22]

*Device Fabrication:* ITO-covered glass substrates (VisionTek Systems) were used as an anode for OLED devices. PEDOT:PSS (4083Ai, Heraeus) was spin coated at 3000 rpm for 1 min, resulting in a 40 nm thick film that was then annealed for 10 min at 130 °C. The subsequent layers were evaporated under vacuum. The NPB hole transport layer of 20 nm thickness was followed by 10 nm mCP electron and exciton blocking layer. The emitting molecule (SBABz4) was co-evaporated with the DPEPO matrix with a weight ratio of 10:90 (SBABz4:DPEPO, 20 nm). On top of that, the hole blocking layers 10 nm DPEPO and 50 nm TPBi were deposited. The device was completed by the top electrode (Ca, 5 nm/Al, 120 nm). The evaporation temperature of SBABz4 was 195 °C to reach 0.3 Å $s^{-1}$.

*EL Measurements: J-V-EL* characteristics were measured with a parameter analyzer (Agilent 4155C) and transient EL with a digitizer card (GaGe CompuScope). The device was driven by a voltage pulse of 1 ms duration and 10 V amplitude. The EL intensity was detected with a silicon photodiode (Hamamatsu S2281) and the EL spectrum with a Photon Control SPM002. The Temperature was controlled by a peltier cooler (Peltron PRG H100), with a temperature range of 244 K to 309 K. Measurements were done under constant nitrogen flow.

*PL Measurements:* The PL samples were evaporated on glass substrates with a weight ratio of 10:90 (SBABz4:DPEPO, 20 nm). Time-resolved and steady-state PL were detected with a calibrated fluorescence spectrometer (Edinburgh Instruments FLS980). Three excitation sources were used: continuous broad-spectrum xenon lamp Xe1, microsecond xenon flash lamp μF920 and an 80 ps pulsed UV diode laser EPL with 375 nm. Temperature was controlled in a cryostat (Janis ST-100) under active vacuum ($<10^{-5}$ mbar).

*Transient and angular-dependent PL Measurements:* The samples were excited normal to the substrate with a He-Cd-Laser (Kimmon HeCd IK5451R) at a wavelength of 325 nm. Spectra for each angle from -80 ° to 80 ° of rotation were taken in 1 ° steps using a liquid-nitrogen cooled CCD detector array attached to a spectrometer (Princeton Instruments Acton SP2300i, Princeton Instruments PyLoN). The polarization of the measured spectra was controlled using a polarizer attached to a rotation mount. The samples were fixed by index-matching fluid to a half-cylindrical prism consisting of BK7 glass. The prompt fluorescence lifetime for the extraction of the intrinsic quantum efficiency was measured using a streak camera system with a spectrograph (Hamamatsu C5680, Princeton Instruments Acton SpectraPro 2300i) and a pulsed Nitrogen-laser at a wavelength of 337 nm (LTB MNL 202-C). The acquired data was averaged over the whole spectrum of the prompt fluorescence and a mono-exponential decay was fitted to extract the lifetime.

AUTHOR INFORMATION

**Corresponding Authors**

*A.S.: e-mail, sperlich@physik.uni-wuerzburg.de.

*M.K.N.: e-mail, mdkhaja.nazeeruddin@epfl.ch.

**ORCID**

Sebastian Weissenseel: 0000-0001-9811-1005

Liudmila G. Kudriashova: 0000-0002-3793-5497

Andreas Sperlich: 0000-0002-0850-6757

Wolfgang Brütting: 0000-0001-9895-8281

Mohammad Khaja Nazeeruddin: 0000-0001-5955-4786

Vladimir Dyakonov: 0000-0001-8725-9573

**Author Contributions**

S.W and N.A.D. contributed equally.

**Funding Sources**

S.W. acknowledges funding from DFG FOR1809. L.G.K., A.S., and V.D. acknowledge EU H2020 for funding through the Grant SEPOMO (Marie Skłodowska-Curie Grant Agreement 722651).

M.S., T.M. and W.B. acknowledge funding from DFG Br1728/20-1 and SolTech.

**Notes**

The authors declare no competing financial interest.

ASSOCIATED CONTENT

**Supporting Information**.

Computational details, gas phase computations, determination of intrinsic quantum efficiency, device characteristics, rate equations of three level system, molecular dynamics simulations.

**Table of Contents Graphic**

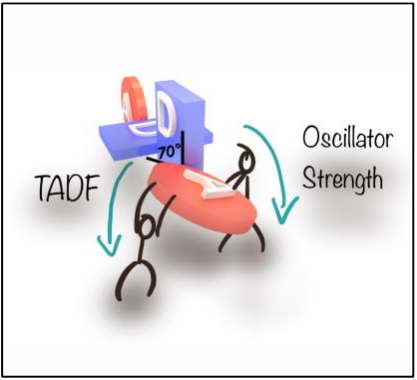

ToC Figure

Supporting Information

# Getting the Right Twist: Influence of Donor–Acceptor Dihedral Angle on Exciton Kinetics and Singlet–Triplet Gap in Deep Blue Thermally Activated Delayed Fluorescence Emitter

*Sebastian Weissenseel, Nikita A. Drigo, Liudmila G. Kudriashova, Markus Schmid, Thomas Morgenstern, Kun-Han Lin, Antonio Prlj, Clémence Corminboeuf, Andreas Sperlich\*, Wolfgang Brütting, Mohammad Khaja Nazeeruddin\*, Vladimir Dyakonov*

**Computational Details**

Gas-phase computations

The ground state geometry of SBABz4 is optimized utilizing density functional theory (DFT) with B3LYP density functional, D3BJ dispersion correction and 6-31G(d,p) basis set.[1–3] For the description of excited states, we employ linear-response time-dependent density functional theory with Tamm-Dancoff approximation (TDA/TD-DFT) along with the state-of-the-art γ-tuned LC-BLYP functional with 6-31G(d,p) basis set to compute the excitation energies and oscillator strengths.[4] The range separation parameter γ was tuned to minimize the difference between minus highest occupied molecular orbital (HOMO) energy and vertical ionization potential. The value of $\gamma$ is 0.165 Bohr$^{-1}$ for SBABz4. In order to validate the TD-DFT results, we also compute excitation energy using ADC(2) method with def2-SVPD basis set, which apart from the small energy shift gives qualitatively similar results as TD-DFT.[5] The characters of different excited states are determined by performing natural transition orbital (NTO) analysis.[6] All computations are done in gas phase. DFT and TDA/TD-DFT computations are performed using Gaussian16 and the ADC(2) computations are performed employing Turbomole 7.1.1.[7,8]

**Table S1.** The excitation energies (E) and oscillator strengths (f) of SBABz4 employing TDA/TD-LC-BLYP*/6-31G(d,p). CT and LE denote charge-transfer and localized excitations, respectively. The values of E and f in parenthesis are computed using ADC(2)/def2-SVPD.

|          | E [eV]       | f           |
|----------|--------------|-------------|
| S1 (CT)  | 3.11 (3.45)  | 0.00 (0.00) |
| S2 (CT)  |              |             |
| T1 (CT)  | 3.10 (3.45)  | 0.00 (0.00) |
| T2 (CT)  |              |             |
| T3 (LE)  | 3.40 (3.54)  | 0.00 (0.00) |

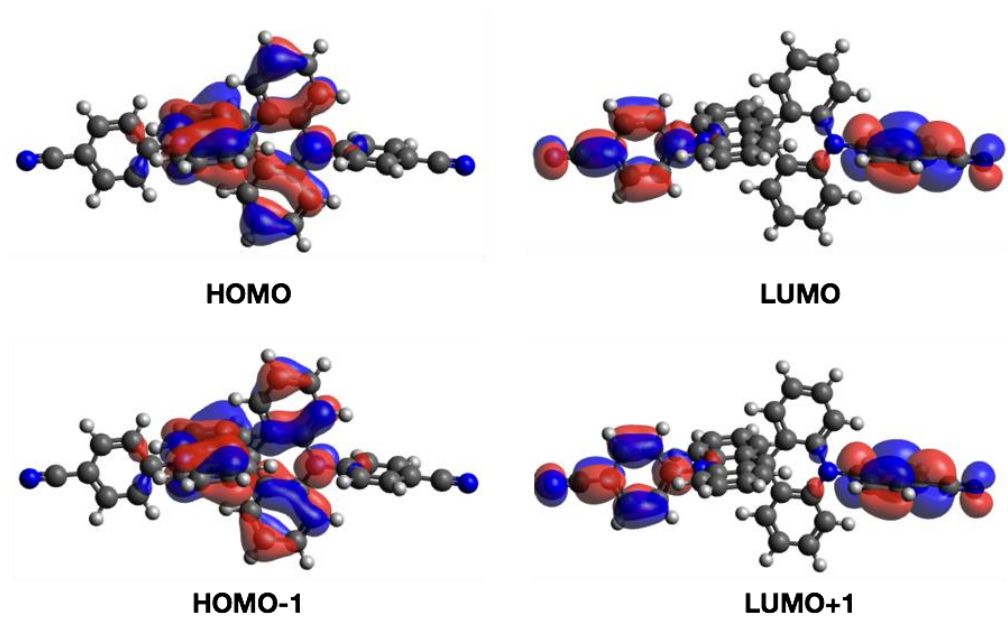

**HOMO**  **LUMO**

**HOMO-1**  **LUMO+1**

**Figure S1.** Nearly degenerate Kohn-Sham HOMOs and LUMOs of **SBABz4** (iso values for the orbitals are 0.02 Bohr$_{-3}$) computed with LC-BLYP*/6-31G(d,p)

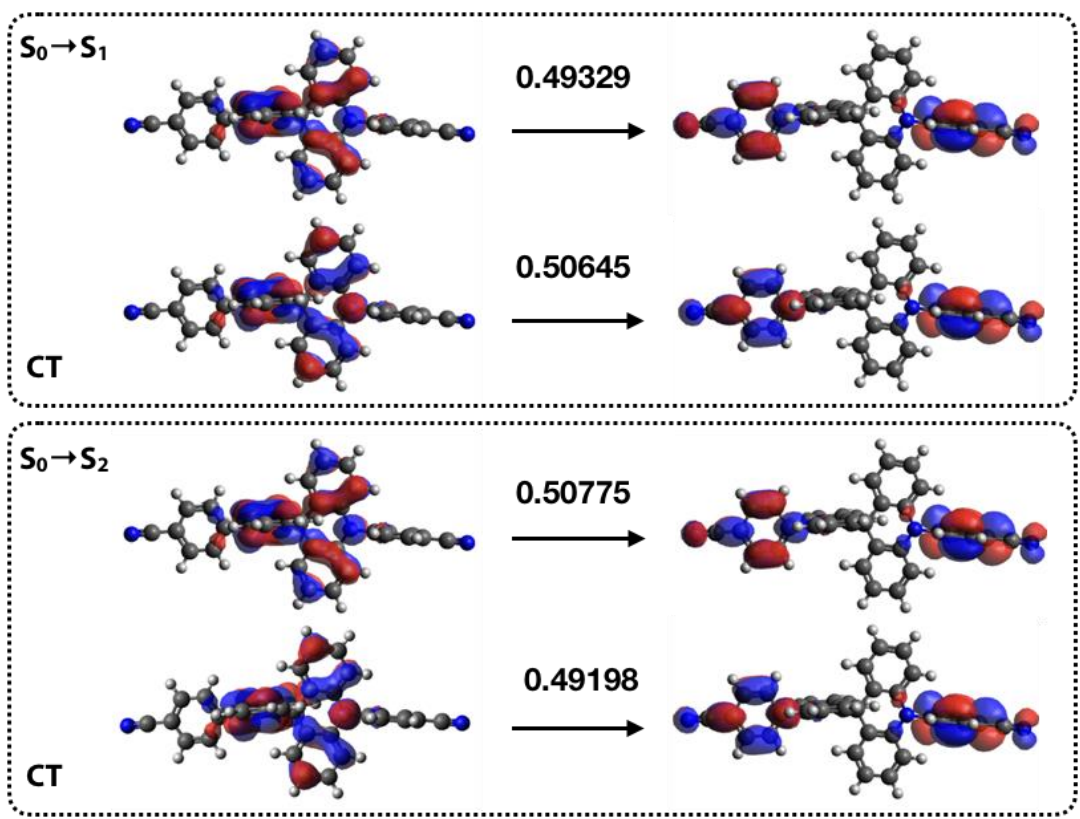

**Figure S2.** Natural transition orbitals (NTOs) of degenerate charge-transfer (CT) singlet states $S_1$ and $S_2$ of **SBABz4** (iso. values for the orbitals are 0.02 Bohr$^{-3}$). LC-BLYP*/6-31G(d,p) was used.

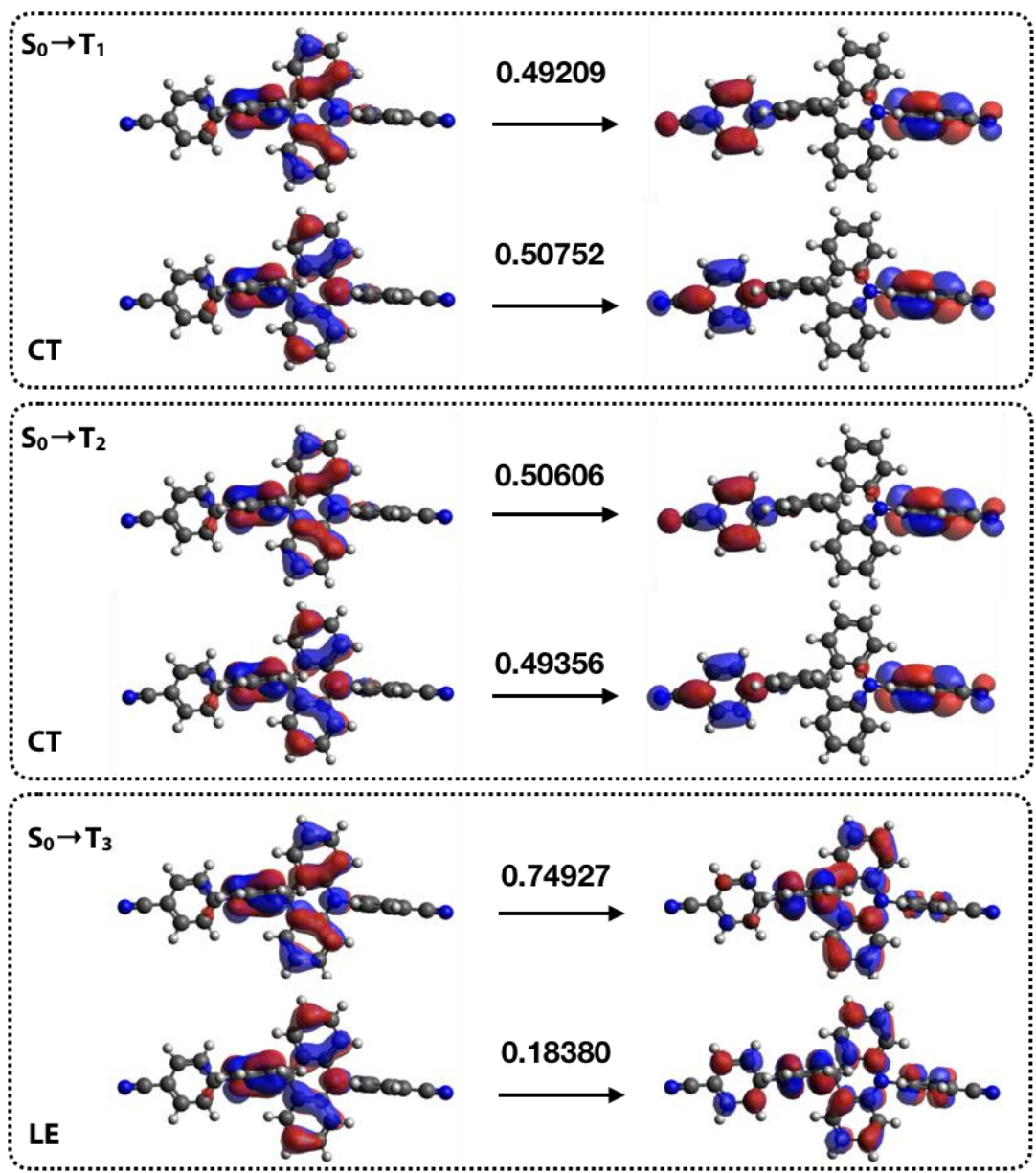

**Figure S3.** Natural transition orbitals (NTOs) of degenerate charge-transfer (CT) triplet states T1, T2 and locally-excited (LE) triplet state T3 of SBABz4 (iso values for the orbitals are 0.02 Bohr$^{-3}$). LC-BLYP*/6-31G(d,p) was used.

Molecular Dynamics Simulation

To take into account the disorder effect present in the real thin film, we performed molecular dynamics (MD) simulations mimicking the physical vapor deposition process similar to the previous computational protocols.[9,10] In every 250 ps, one molecule (either DPEPO or SBABz4) was deposited, where the probability of choosing host or OLED molecules depends on the final ratio (in this case SBABz4 is 10 mol%). We finally reached a ratio of SBABz4:DPEPO = 112:888. Since the deposition rate is very high in MD simulations, we kept the whole system at 400 K during the deposition process followed by annealing at 400 K for 10 ns, allowing deposited molecules to find local minimum during the process. The whole system was then equilibrated at 300 K for 20 ns and a 10 ns production run was performed to generate 1000 MD snapshots for the dihedral angle analysis. The dihedral angle distribution computed using 1 MD snapshot (112 molecules) is similar to that obtained from 1000 snapshots (112000 molecules), as shown in Figure S4. Therefore, the average $\Delta E_{ST}$ computed with TDA/TD-DFT was evaluated using all the SBABz4 molecules in one MD snapshot. To account for the effect of the surrounding DPEPO host molecules on the excited states properties, we adopted a predictive and efficient method proposed by Brédas *et al.*, involving a polarizable continuum model (PCM) and TDA/TDDFT with a γ-tuned range-separated functional. [11] The dielectric constant was chosen to be 3, which is a fair estimation for organic solids. Brédas *et al.* also showed that the γ value would be lower in the presence of dielectrics and needs to be reoptimized. [12] Therefore, we retuned the γ for LC-BLYP functional, which is 0.051 Bohr-1. The MD simulations were performed under periodic boundary conditions with GROMACS package and AMBER force field, where the polarization effect if not taken into account.[13–15] All bonding parameters (bonds, angles and dihedrals) were optimized to reproduce the potential energy of reference level (B3LYP-D3BJ/6-31G(d,p)) using ffTK toolkit and Gaussian16.[16] Atomic charges were obtained by the restrained electrostatic potential (RESP) procedure based on B3LYP/6-31G(d,p).[17] The temperature control was accomplished using velocity rescaling with a stochastic term ($\tau_T$ = 1.0 ps).[18] The time step used in all simulations was 1 fs, and bonds involving H atoms were constrained using the Linear Constraint Solver (LINCS) algorithm. A cutoff of 12 Å was applied to the van der Waals interaction through force-switch mode. As for electrostatic interactions, the particle mesh Ewald (PME) method was employed with a 0.12 nm Fourier spacing.

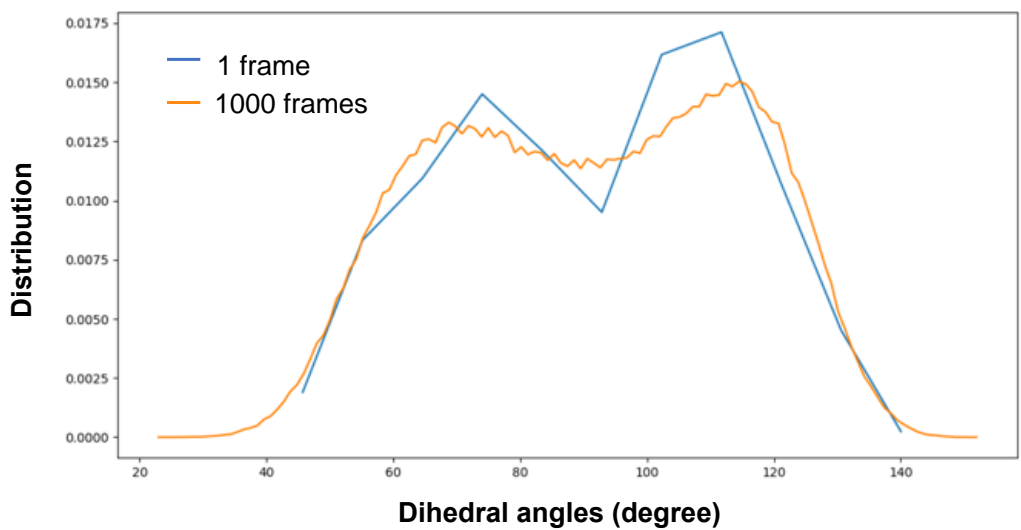

**Figure S4.** Dihedral angle distribution calculated from 1 MD snapshot and 1000 snapshots.

Dihedral angle of SBABz4

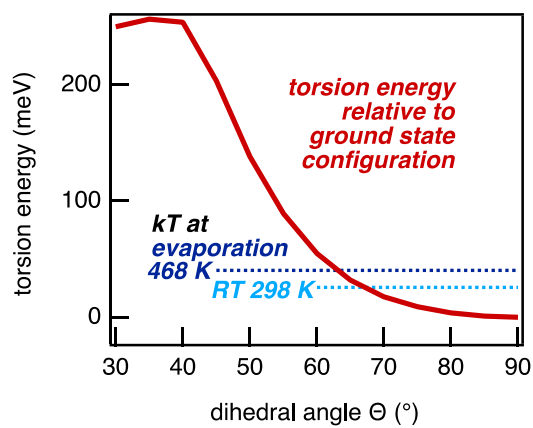

**Figure S5.** Relative energy of SBABz4 with different dihedral angles (benzene-acridine) in constrained-relaxation dihedral scan

Determination of the intrinsic quantum efficiency

The intrinsic quantum efficiency *IQE* of an emitter can be determined utilizing the Purcell effect in a microcavity and transient photoluminescence measurements.[19] The lifetime $\tau$ of the emitting species is modified by the Purcell effect according to

$$\tau = \frac{1}{F \cdot IQE + (1 - IQE)} \cdot \tau_0$$

where $F$ and $\tau_0$ describe the Purcell factor of the cavity and the intrinsic excitonic lifetime in the absence of a cavity, respectively. A significant variation of the microcavity is achievable by placing the emitting dipole near a reflecting silver layer and varying the distance between the dye and the mirror as shown in the inset in Figure S6. The Purcell factor of the microcavity can be calculated with the same optical simulation that is also usable for the calculation of the maximum EQE. Here we take the prompt fluorescence lifetime for SBABz4 doped in DPEPO at a concentration of 10 %. Modifications of the microcavity were realized by evaporating an additional spacer layer of 1,4-bis(triphenylsilyl)benzene (UGH-2) with varying thickness covered with a reflective silver layer of at least 100 nm. Figure S6 shows the extracted lifetimes together with calculated behavior of $\tau$ in dependence of the spacer layer thickness. Fitting the expression above yields the intrinsic radiative quantum efficiency and the intrinsic fluorescence lifetime of 0.36 ± 0.07 and 11.3 ns, respectively. The determined radiative quantum yield is in good agreement with photoluminescence quantum yield measurements published previously.[20]

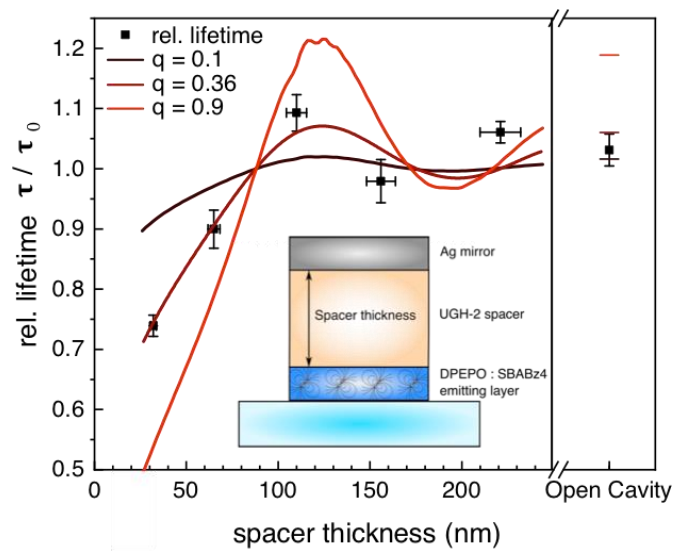

**Figure S6.** Experimental and calculated values of the prompt fluorescence lifetime in dependence of the UGH-2 spacer layer thickness and sketch of the sample structure. For clarity the lifetimes were normalized to the intrinsic value of $\tau_0$ = 11.3 ns. Note that open cavity means that only the emitter layer but no additional spacer and silver were evaporated on the sample.

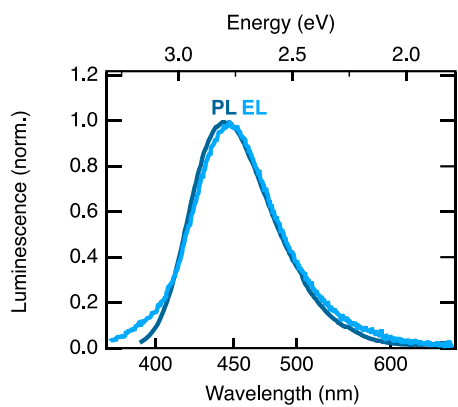

**Figure S7.** Normalized PL and EL spectra of SBABz4. PL was measured on a film with 5wt% SBABz4:DPEPO, while the EL was detected from a fully processed OLED device.[20]

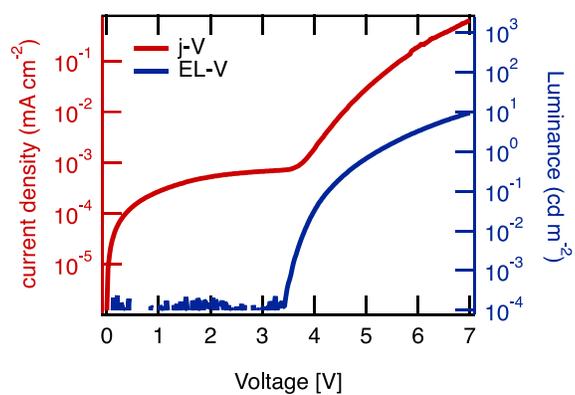

**Figure S8.** J-V characteristics (red) and EL-V-characteristics (blue) of a SBABz4:DPEPO OLED

Analysis of transient PL and EL

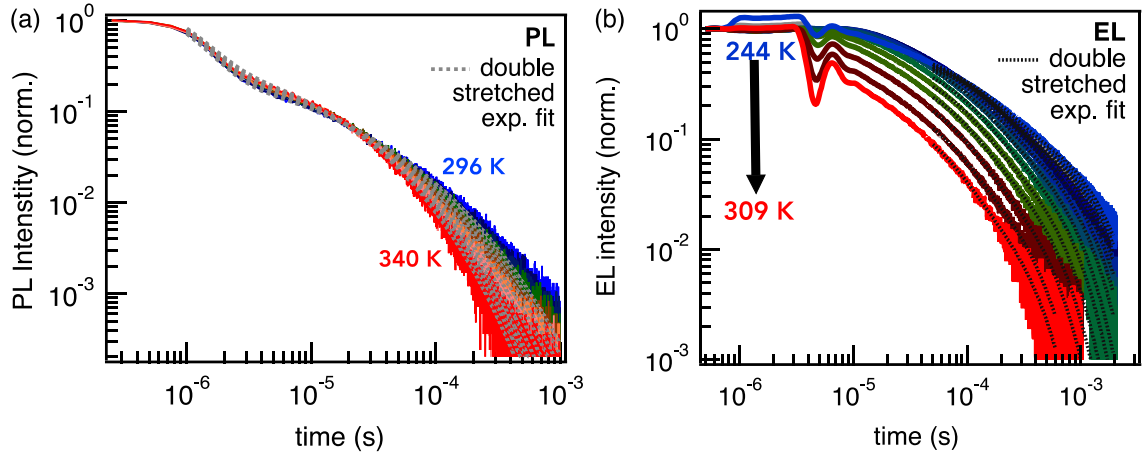

**Figure S9.** (a) Temperature dependent trPL and (b) trEL

rate equations for a three level system:

$$\frac{dn}{dt} = -(k_{10} + q_{10} + k_{13}) \cdot n + k_{31} \cdot m$$

$$\frac{dm}{dt} = k_{13} \cdot n - (k_{30} + q_{30} + k_{31}) \cdot m$$

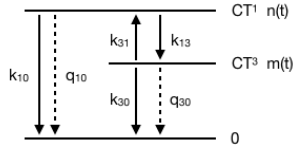

With:

$$\begin{pmatrix} a & b \\ c & d \end{pmatrix} \equiv \begin{pmatrix} k_{10} + q_{10} + k_{10} & k_{31} \\ k_{13} & k_{30} + q_{30} + k_{31} \end{pmatrix}$$

The characteristic equation for this system is

$$\lambda_{1,2} = \frac{1}{2}\left(a + d \pm \sqrt{(a+d)^2 - 4(ad - bc)}\right) = \frac{1}{2}(a + d \pm \xi)$$

Where $\xi \equiv \sqrt{(a+d)^2 - 4(ad - bc)}$

The exact solution for this system is

$$n(t) = \frac{1}{\xi}[(n_0(\lambda_1 - d) - bm_0) \cdot e^{-\lambda_1 t} + (n_0(d - \lambda_2) + bm_0) \cdot e^{-\lambda_2 t}]$$

$$m(t) = \frac{1}{\xi}[(m_0(d - \lambda_2) - cn_0) \cdot e^{-\lambda_1 t} + (n_0(\lambda_1 - d) + cn_0) \cdot e^{-\lambda_2 t}]$$

Rate of photon emission from the singlet level n

$$\frac{d\nu_{em}}{dt} = k_{10}n(t) \longrightarrow \nu_{em} = \int_0^\infty \frac{d\nu_{em}}{dt} dt = \int_0^\infty k_{10}n(t)dt$$

external electroluminescence quantum efficiency (EQE) is defined as the ratio between the number of photons emitted by the device and the total number of charge carriers:

$$\Phi_{EL} = \frac{\nu_{em}}{n_0 + m_0}$$

The expression for EQE can then be written as

$$\Phi_{EL} = \frac{k_{10}}{n_0 + m_0} \frac{1}{\xi} \left( \frac{n_0(\lambda_1 - d) - bm_0}{\lambda_1} + \frac{n_0(d - \lambda_2) + bm_0}{\lambda_2} \right)$$

$$\Phi_{EL} = \frac{k_{10}}{n_0 + m_0} \frac{1}{\xi} (n_0 d + m_0 b) \left( \frac{1}{\lambda_2} - \frac{1}{\lambda_1} \right)$$

With prompt component

$$\Phi_{EL,pr} = \frac{k_{10}}{n_0 + m_0} \frac{1}{\xi} \frac{n_0(\lambda_1 - d) - bm_0}{\lambda_1}$$

and delayed component

$$\Phi_{EL,TADF} = \frac{k_{10}}{n_0 + m_0} \frac{1}{\xi} \frac{n_0(d - \lambda_2) + bm_0}{\lambda_2}$$

The ratio between prompt and delayed components results in

$$\frac{\Phi_{pr}}{\Phi_{TADF}} = \frac{n_0(\lambda_1 - d) - m_0 b}{\lambda_1} \cdot \frac{\lambda_2}{n_0(d - \lambda_2) + bm_0} = \frac{\lambda_2}{\lambda_1} \frac{0.25(\lambda_1 - d) - 0.75b}{0.25(d - \lambda_2) + 0.75b} \quad \text{with} \quad \begin{matrix} n_0 = 0.25N \\ m_0 = 0.75N \end{matrix}$$

**Approximations made from physical properties known from PLQY measurements (SI in previous publication):** [20]

$q_{10} = 0$

(only radiative decay from singlet, since the prompt PL component is temperature-independent, Supporting Information in previous publication Figure S10a [20])

$k_{30} = 0$

(phosphorescence was not observed at room temperature);

$k_{10} \approx k_{13}$

(the rate constant of radiative decay from the singlet is close to the ISC rate constant, since the triplet is effectively populated);

$k_{31} \approx q_{30}$

(rate constant of RISC is in the order of triplet non-radiative rate constant, since both processes are observed);

$k_{10} \gg k_{31}$

(rate constant of singlet radiative decay significantly exceeds RISC, since the observed prompt decay is way faster than TADF).

$$\frac{\Phi_{pr}}{\Phi_{TADF}} \approx \frac{\lambda_2}{k_{31}} \frac{1}{3 + \frac{k_{13}}{k_{10}+k_{13}}} \quad \text{while} \quad \lambda_1 - d \approx \lambda_1 \approx a \quad \text{and} \quad d - \lambda_2 \approx \frac{bc}{a}$$

The reverse intersystem crossing can then be written as:

$$k_{RISC} \equiv k_{31} \approx \lambda_2 \cdot \frac{\Phi_{TADF}}{\Phi_{pr}} \cdot \frac{1}{3 + \frac{k_{13}}{k_{10}+k_{13}}}$$

$$ln(k_{RISC}) \approx ln(\lambda_2 \cdot \Phi_{TADF}) + const.$$

Also known:

$$k_{RISC} \propto exp\left(\frac{-\Delta E_{ST}}{k_b T}\right)$$

evaluation formula to determine the activation energy from trEL:

$$ln(\lambda_2 \cdot \Phi_{TADF}) + const. \mathrel{\hat=} \frac{\Delta E_{ST}}{k_b}\frac{1}{T} + const.$$

SUPPORTING INFORMATION REFERENCES